\documentclass[twocolumn,showpacs,superscriptaddress]{revtex4}
\usepackage{amsfonts}
\usepackage{amsmath}
\usepackage{amssymb}
\usepackage{graphicx}%
\providecommand{\U}[1]{\protect\rule{.1in}{.1in}}

\begin{document}
\title{First-principles study of pressure-induced phase transition and electronic property of PbCrO$_{3}$}
\author{Bao-Tian Wang}
\affiliation{Institute of Theoretical Physics and Department of Physics, Shanxi University,
Taiyuan 030006, People's Republic of China}
\affiliation{State Key Laboratory of Magnetism, Beijing National Laboratory for Condensed
Matter Physics, Institute of Physics, Chinese Academy of Sciences, Beijing
100080, People's Republic of China}
\author{Wen Yin}
\affiliation{State Key Laboratory of Magnetism, Beijing National Laboratory for Condensed
Matter Physics, Institute of Physics, Chinese Academy of Sciences, Beijing
100080, People's Republic of China}
\author{Wei-Dong Li}
\affiliation{Institute of Theoretical Physics and Department of Physics, Shanxi University,
Taiyuan 030006, People's Republic of China}
\author{Fangwei Wang}
\thanks{Author to whom correspondence should be addressed; fwwang@aphy.iphy.ac.cn}
\affiliation{State Key Laboratory of Magnetism, Beijing National Laboratory for Condensed
Matter Physics, Institute of Physics, Chinese Academy of Sciences, Beijing
100080, People's Republic of China}

\pacs{61.50.Ah, 71.15.Mb, 75.25.+z}

\begin{abstract}
We have performed a systematic first-principles investigation to
calculate the structural, electronic, and magnetic properties of
PbCrO$_{3}$, CrPbO$_{3}$ as well as their equiproportional
combination. The local density approximation (LDA)$+U$ and the
generalized gradient approximation$+U$ theoretical formalisms have
been used to account for the strong on-site Coulomb repulsion among
the localized Cr $3d$ electrons. By choosing the Hubbard \emph{U}
parameter around 4 eV within LDA$+U$ approach, ferromagnetic, and/or
antiferromagnetic ground states can be achieved and our calculated
volumes, bulk moduli, and equation of states for PCO-CPO in $R3$
phase and PCO in $Pm\bar{3}m$ or $R3c$ phases are in good agreement
with recent experimental Phase I and Phase II [W. Xiao \emph{et
al.}, PNAS \textbf{107}, 14026 (2010)], respectively. Under
pressure, phase transitions of $R3$ PCO-CPO to $Pm\bar{3}m$ PCO at
1.5 GPa and $R3$ PCO-CPO to $R3c$ PCO at -6.7 GPa have been
predicted. The abnormally large volume and compressibility of Phase
I is due to the content of CrPbO$_{3}$ in the experimental sample
and the transition of PbO$_{6/2}$ octahedron to CrO$_{6/2}$ upon
compression. Our electronic structure study showed that there will
occur an insulator-metal transition upon the phase transitions.
Clear hybridization of Cr 3\emph{d} and O 2\emph{p} orbitals in wide
energy range has been observed.
\end{abstract}
\maketitle

\section{INTRODUCTION}
Strongly correlated electron systems of transition-metal oxides with
ABO$_{3}$ cubic perovskite or pseudo cubic perovskite structures
exhibit particular interesting physical properties.
\cite{Zhong,Imada,Tokura,Dagotto} Their ferroelectric,
ferromagnetic, ferroelastic, multiferroic, and/or magnetoresistive
features originate from the mutual interplay of various degrees of
freedom, including lattice, spin, charge and orbital, in their
partially filled B site 3\emph{d} electrons. The multiple chemical
characters of the A ion with lone pair electrons, especially for
Bi$^{3+}$ and Pb$^{2+}$, also play an important role.
\cite{Neaton,Picozzi} Correctly describing of their electronic and
magnetic structures are critical.

In late 1960s, a few groups
\cite{Roth1,Roth2,Chamberland1,Chamberland2,Goodenough,Weiher}
successfully synthesized some perovskites containing Cr$^{4+}$
(CaCrO$_{3}$, SrCrO$_{3}$, and PbCrO$_{3}$) under high temperature
and high pressure. For PbCrO$_{3}$ (PCO), a lattice constant of
about 4.00 \AA\ for the cubic structure was determined by X-ray
diffraction on single crystal and powder samples and powder neutron
diffraction \cite{Roth1,Roth2,Chamberland1}. It was reported that
the PCO is an antiferromagnetic (AFM) G-type semiconductor with a
magnetic moment of $\mathtt{\sim}$1.9 $\mu_{B}$ on each Cr ion
\cite{Roth1,Roth2} and with 0.27 eV activation energy.
\cite{Chamberland1} The Neel temperature ($T_{\text{N}}$) of about
240 K was obtained by examining the magnetic and transport
properties. \cite{Roth1,Roth2} For CaCrO$_{3}$ and SrCrO$_{3}$,
preliminary structural, magnetic, and conductive properties were
also investigated. \cite{Chamberland2,Goodenough,Weiher} From then
on, in a long period of more than thirty years little works had
focused on these systems due to difficulty of synthesis. However, a
renewed interest on them has been inspired on their transport and
magnetic properties, insulator-metal transition, and pressure
behaviors.
\cite{Zhou,Ortega,Komarek,ArevaloJSSC,ArevaloJPCM,ArevaloInorg,Xiao}
Anomalous properties of Seebeck coefficient, thermal conductivity,
magnetic susceptibility, and room-temperature compressibility have
been observed for SrCrO$_{3}$ by Zhou \emph{et al.} They concluded
that SrCrO$_{3}$ is nonmagnetic (NM) insulator and CaCrO$_{3}$ is
also an insulator at low temperature. But more recent studies
claimed that both these perovskites are AFM metals.
\cite{Ortega,Komarek} An orbital ordering transition from
$t_{2g}^{2}$ to $d_{xy}^{1}(d_{xz}d_{yz})^{1}$ and electronic phase
coexistence of $C$-AFM tetragonal and NM cubic phases have been
discovered in SrCrO$_{3}$. \cite{Ortega} Komarek \emph{et al.}
reported that CaCrO$_{3}$ is an intermediately correlated metal with
similar $C$-type AFM ground state. Thus, whether these systems are
metallic, strongly correlated, and spin ordered is still
controversial. \cite{Streltsov,Lee}

Recent experimental works of PCO concentrated on its structure,
electron energy loss spectroscopy, magnetic structure, and high
pressure phase transition.
\cite{ArevaloJSSC,ArevaloJPCM,ArevaloInorg,Xiao} Electron
diffraction and high-resolution electron microscopy study revealed
that the microstructure of ``PbCrO$_{3}$" is a rather complex
perovskite with a compositionally modulated
$a_{p}\times3a_{p}\times(14\mathtt{\sim}18)a_{p}$ superlattice
structure, where $a_{p}=4.002$ \AA\ is the lattice constant of the
cubic PCO perovskite. \cite{ArevaloJSSC} The magnetic structure of
PCO is also complex. Alario-Franco \emph{et al.} reported AFM
ordering of the chromium moments at $T_{\text{N}}$$\mathtt{\sim}$245
K with a spin-reorientation at temperature range of 185 K to 62 K
and their magnetic hysteresis loops for ``PbCrO$_{3}$" suggested
weak ferromagnetism at low temperature. \cite{ArevaloInorg} Their
resistivity measurements indicated two activation energies ranges
with 0.11 and 0.26 eV in different temperatures. As for pressure
study, recent work performed by Xiao \emph{et al.} \cite{Xiao}
observed a large volume collapse in the isostructural transition of
cubic PCO perovskite at $\mathtt{\sim}$1.6 GPa from Phase I to Phase
II. They concluded that the transition seems not related with any
change of electronic state, but probably has tight relation with the
abnormally large volume and compressibility of the Phase I. The real
Phase I might be a kind of mixture of PbCrO$_{3}$-CrPbO$_{3}$
(PCO-CPO) combination due to the fact that the cubic lattice
constant is enlarged if the CrO$_{6/2}$ octahedron could be replaced
by PbO$_{6/2}$ \cite{Xiao}.

In present study, we focus our sight on PCO, PCO-CPO, and
CrPbO$_{3}$ (CPO) in the cubic perovskite structure (space group
$Pm\bar{3}m$) and some possible distorted perovskite structures,
such as $R3c$, $R3$, and $P4/mmm$ phases. Electronic and magnetic
properties as well as pressure behaviors have been systematically
investigated by the first-principles electronic structure
calculations based on density functional theory (DFT) and
DFT+\emph{U} schemes due to Dudarev \emph{et al}. \cite{Dudarev} The
validity of the ground-state calculation is carefully tested. Our
calculated lattice parameter and bulk modulus \emph{B} for cubic PCO
are well consistent with previous local density approximation (LDA)
and generalized gradient approximation (GGA) results \cite{Xiao}.
The total energy, lattice constant, bulk modulus \emph{B}, and spin
moment of Cr ion for NM, ferromagnetic (FM), and AFM phases
calculated in wide range of effective Hubbard \emph{U} parameter are
presented and our calculated results within LDA+\emph{U} with
\emph{U}=3-4 eV for PCO in $Pm\bar{3}m$ or $R3c$ phases and PCO-CPO
in $R3$ phase accord well with experimental \cite{Xiao} Phase II and
Phase I, respectively. Our calculated spin moment by LDA+\emph{U} is
in good agreement with recent experimental value of saturation
moment $M_{\rm{sat}}$=1.70 $\mu_{B}$, which is deduced from the
effective moment $M_{\rm{eff}}$=2.51 $\mu_{B}$ \cite{ArevaloInorg}
according to the relation
$M_{\rm{eff}}$=2$[S(S+1)]^{1/2}$=$[M_{\rm{sat}}(M_{\rm{sat}}+2)]^{1/2}$.
The P-V relations of PCO and PCO-CPO are calculated to compare with
experiment. The insulting property of ``PbCrO$_{3}$" at ambient
condition is successfully predicted.

\section{computational methods}

First-principles DFT calculations on the basis of the frozen-core
projected augmented wave (PAW) method of Bl\"{o}chl \cite{PAW} are
performed within the Vienna \textit{ab initio} simulation package
(VASP) \cite{Kresse3}, where the exchange and correlation effects
are described by the DFT within LDA and GGA \cite{LDA,GGA}. For the
plane-wave set, a cutoff energy of 500 eV is used. The
\emph{k}-point meshes in the full wedge of the Brillouin zone (BZ)
are sampled by 12$\times$12$\times$12 and 6$\times$6$\times$6 grids
according to the Monkhorst-Pack \cite{Monk} scheme for PCO, CPO, and
PCO-CPO in their cubic and rhombohedral unit cells, respectively.
The cubic perovskite structure are built for nonmagnetic (NM) and
ferromagnetic (FM) calculations and the rhombohedral unit cells
(Fig. 1) for G-type antiferromagnetic (AFM) configuration, which is
($\frac{1}{2}$, $\frac{1}{2}$, $\frac{1}{2}$) order in terms of the
original perovskite cell. The rhombohedral unit cell for PCO in its
cubic structure is constructed using the $R3c$ space group
($\alpha$=60$^{\circ}$) with Pb atoms in $2a$(0, 0, 0) site, Cr in
$2a$($\frac{1}{4}$, $\frac{1}{4}$, $\frac{1}{4}$) site, and O in
$6b$($\frac{1}{2}$, 0, $\frac{1}{2}$) site. Moving away slightly
atoms from these high symmetry positions will lower the symmetry and
consequently build the cell in real $R3c$ or $R3$ phase. In this
study, the Pb
5$d$$^{10}$6$s$$^{2}$6$p$$^{2}$, Cr 3$d$$^{5}$4$s$$^{1}$, and O 2$s$$^{2}$%
2$p$$^{4}$ orbitals are explicitly included as valence electrons. The strong
on-site Coulomb repulsion among the localized Cr 3\emph{d} electrons is
described by using the formalism formulated by Dudarev \emph{et al.}
\cite{Dudarev}. In this scheme, the total LDA (GGA) energy functional is of
the form
\begin{align}
E_{\mathrm{{LDA(GGA)}+U}}  &  =E_{\mathrm{{LDA(GGA)}}}\nonumber\\
&  +\frac{U-J}{2}\sum_{\sigma}[\mathrm{{Tr}\rho^{\sigma}-{Tr}(\rho^{\sigma
}\rho^{\sigma})],}%
\end{align}
where $\rho^{\sigma}$ is the density matrix of \emph{d} states with
spin $\sigma$, while \emph{U} and \emph{J} are the spherically
averaged screened Coulomb energy and the exchange energy,
respectively. In this work, the Coulomb \emph{U} is treated as one
variable, while the parameter \emph{J} is set to 0.5 eV. Since only
the difference between \emph{U} and \emph{J} is meaningful in
Dudarev's approach, therefore, we label them as one single parameter
\emph{U} for simplicity. To obtain the energy data under different
pressures, we perform the structure-relaxation calculations at a
series of fixed volumes. The corresponding pressure values are
deduced from the energy-volume data by
$P$=$-\partial$$E$/$\partial$$V$.

\section{results}

\subsection{Structural and magnetic properties of cubic PCO}
Both spin-unpolarized and spin-polarized calculations are performed
for cubic PCO. For NM, FM, and AFM configurations, the total
energies (-39.685 eV, -39.706 eV, and -39.818 eV, respectively)
calculated within the DFT (\emph{U}=0) have no visible differences.
After turning on the Hubbard \emph{U}, the NM phase is not
energetically favorable both in the LDA+\emph{U} and GGA+\emph{U}
formalisms compared with FM and AFM phases. In the following, we
only present results of FM and AFM configurations. The dependences
of total-energy differences (per formula unit at respective optimum
geometries) between FM and AFM
($E_{\text{FM}}\mathtt{-}E_{\text{AFM}}$), lattice parameter, bulk
modulus, and spin moments of Cr ions on \emph{U} for cubic PCO are
shown in Fig. 2. The theoretical equilibrium volume, bulk modulus
\emph{B}, and pressure derivative of the bulk modulus
\emph{B$^{\prime}$} are obtained by fitting the third-order
Birch-Murnaghan equation of state (EOS) \cite{Birch}. For
comparison, recent experimental values \cite{Xiao} of $a_{0}$ and
\emph{B} for Phase I and Phase II as well as the experimental
results of spin magnetic moment of Cr ions are also shown. In LDA or
GGA (\emph{U}=0 eV), the total energy of AFM phase is lower than
that of FM phase. However, as shown in Fig. 2(a), the total energy
of the FM phase decreases to become lower than that of the AFM phase
when increasing \emph{U}. At a typical value of \emph{U}=4 eV, the
total-energy differences between FM and AFM
($E_{\text{FM}}\mathtt{-}E_{\text{AFM}}$) reach their minimums of
$-$74 meV and $-$91 meV within the LDA+\emph{U} and GGA+\emph{U}
formalisms, respectively. Note that we have also considered other
types of AFM configurations and our results show that the
\emph{G}-AFM in ($\frac{1}{2}$, $\frac{1}{2}$, $\frac{1}{2}$) order
is the most stable state. These results are consistent with recent
experimental observations \cite{ArevaloInorg}, where they reported
the AFM ordering of the chromium moments at
$T_{\text{N}}$$\mathtt{\sim}$245 K with a spin-reorientation at
temperature range of 185 K to 62 K and their magnetic hysteresis
loops for ``PbCrO$_{3}$" suggested weak ferromagnetism at low
temperature. As clearly shown in Fig. 2(b), our calculated lattice
parameters of cubic PCO within two DFT+\emph{U} approaches are all
by large smaller than the experimental value $a_{0}$=4.013 \AA \
\cite{Roth1,Roth2,Chamberland1,Xiao} of Phase I. Corresponding
results of bulk modulus are dramatically bigger than that obtained
by experiment (\emph{B}=59 GPa) for Phase I [Fig. 2(c)]. But the
experimental lattice parameter $a_{0}$=3.862 \AA \ and bulk modulus
\emph{B}=187 GPa of Phase II well lies in the range of our
calculated results of cubic PCO, which supports their conclusion of
cubic PCO perovskite \cite{Xiao} that the real Phase I might be a
kind of mixture of random PCO-CPO combination. In next subsection,
we will discuss carefully the structural properties of PCO-CPO.

As shown in Fig. 2, the tendencies of $a_{0}$, \emph{B}, and spin
moments of Cr ions for FM phase with \emph{U} are similar to that of
the AFM phase. Results of NM phase (not presented) further indicates
its unstable nature at low temperature. As shown in Fig. 2(b), LDA
underestimates the lattice parameter with respect to the
experimental value while GGA coincides well with experiment. Our LDA
and GGA results for NM phase accords well with previous
first-principle calculations \cite{Xiao}. Result of LDA is due to
its typical overbinding character. The LDA+\emph{U} method will lead
to relative larger equilibrium volume compared to the LDA and
therefore improves the agreement with experiment, especially for FM
and AFM phases. At a typical value \emph{U}=4 eV, the LDA+\emph{U}
gives $a_{0}$=3.822 \AA \ for AFM cubic PCO which is very close to
the experimental value. On the other hand, the GGA+\emph{U} enlarges
the underbinding effect with increasing Hubbard \emph{U}. As a
comparison, at \emph{U}=4 eV, the GGA+\emph{U} gives $a_{0}$=3.921
\AA. As for the dependence of bulk modulus \emph{B} on \emph{U}
shown in Fig. 2(c), it is clear that the LDA results (230-153 GPa)
are always higher than the GGA results (187-113 GPa), which is due
to the above mentioned underbinding effect of the GGA approach. At a
typical value \emph{U}=4 eV, the LDA+\emph{U} and GGA+\emph{U} give
\emph{B}=198 (\emph{B$^{\prime}$}=4.6) and 148 GPa
(\emph{B$^{\prime}$}=5.1) for AFM cubic PCO, respectively.
Obviously, result calculated within LDA+\emph{U} with \emph{U}=4 eV
is consistent with the experimental value of \emph{B}=187 GPa
\cite{Xiao}. For the dependence of spin moments of Cr ions on
\emph{U} shown in Fig. 2(d), we see clear increasing amplitude of
magnetic moments with \emph{U} for both FM and AFM phases. Our
calculated value of 1.47 $\mu_{B}$ using LDA for AFM phase is in
good agreement with previous LMTO calculation value of 1.414
$\mu_{B}$ \cite{Jaya}. At a typical value \emph{U}=4 eV, the
LDA+\emph{U} and GGA+\emph{U} give magnetic moments of 2.46 and 2.62
$\mu_{B}$ for AFM PCO, respectively, both of which exceed recent
experimental value of 1.70 $\mu_{B}$ \cite{ArevaloInorg}. Similar
trend has also been exhibited in study of BiFeO$_{3}$ \cite{Neaton}.
Therefore, inclusion of on-site Coulomb energy for adequately
describing the structural and magnetic properties is crucial. In
study of CrO$_{2}$, x-ray absorption and resonant photoemission
spectroscopy support the importance of Coulomb correlations
\cite{Dedkov}. Huang \emph{et al.} concluded that the on-site
Coulomb interaction energy of CrO$_{2}$ is 3-4 eV through comparing
their experimental measurements and LDA+\emph{U} calculations. They
found that the shift of Cr 3\emph{d} spin-up DOS slightly away from
the Fermi level increases the Cr spin moment. In our present study,
similar shift of Cr 3\emph{d} DOS is also observed (see below).
Overall, comparing with the experimental data, the accuracy of our
atomic-structure prediction for AFM cubic PCO is quite satisfactory
by tuning the effective Hubbard parameter \emph{U} in a range of 3-4
eV within the LDA+\emph{U} approach, which supplies the safeguard
for our following study of electronic structure and pressure
behaviors of PCO. In following study, we present our results within
LDA and LDA+\emph{U} with constant \emph{U}=4 eV.

\subsection{Phase transition analysis}
After previous analysis, one question arises: What's the ground
state of experimentally observed ``PbCrO$_{3}$" at ambient
condition. In this section, we will test carefully the ground state
of ``PbCrO$_{3}$" by calculating the total energies of PCO, PCO-CPO,
and CPO in cubic perovskite structure and some possible distorted
perovskite structures. Their relative energies in $G$-AFM
configuration calculated within LDA+$U$ are shown in Fig. 3.
Clearly, although the equilibrium lattice parameter (3.998 \AA) for
$Pm\bar{3}m$ phase of PCO-CPO accords well with experimental data of
Phase I \cite{Xiao}, the $Pm\bar{3}m$ phase of PCO-CPO and CPO are
not energetically favorable in the whole range of volume. After
testing various possible crystal structures and atomic arrangements,
we find that $R3c$ phase of PCO and $R3$ phase of PCO-CPO possess
same energy level with that of cubic PCO in some volumes. In the
whole range of volume, $R3c$ PCO is more energetically favorable
than cubic PCO. Considering this material only can be synthesized
under high temperature and high pressure, the high-pressure phase of
PCO can be understood as crystallizing in $Pm\bar{3}m$ or $R3c$
structure. Increasing the cell volume, as shown in Fig. 3, the $R3$
PCO-CPO becomes more energetically favorable than the two
high-pressure phases. This clearly indicates that the structure of
``PbCrO$_{3}$" at ambient pressure is $R3$ PCO-CPO. Therefore, upon
compression there will occur structural phase transition among the
three phases. As shown in the inset of Fig. 3, phase transitions of
$R3$ PCO-CPO to $Pm\bar{3}m$ PCO at 1.5 GPa and $R3$ PCO-CPO to
$R3c$ PCO at -6.7 GPa can be predicted by the slopes of the common
tangent rule. The former value coincides well with recent
experimentally observed value of 1.6 GPa, while the latter value is
smaller than that. As a result, the crystal structures of PCO under
high pressure need more works to clarify. Our present study give two
most possible structures: $Pm\bar{3}m$ or $R3c$. Concerning the
energetics of the transitions from $R3$ PCO-CPO to $Pm\bar{3}m$ PCO
at 1.5 GPa and from $R3$ PCO-CPO to $R3c$ PCO at -6.7 GPa, Fig. 3
clearly shows that per formula unit of PCO-CPO needs 97 meV and 458
meV of energy, respectively.

In Table I, we present the calculated lattice parameters for PCO in
$R3c$ phase and PCO-CPO in $R3$ structure. It can be found that the
equilibrium volume of $R3c$ PCO is consistent with that of
experimental Phase II and $R3$ PCO-CPO comparable to Phase I
\cite{Xiao}. Although Xiao \emph{et al.} reported that both the
Phase I and Phase II crystallize in cubic perovskite structure, our
calculations are different from their observations. Our present
results need more experimental works to test. For $R3c$ PCO and $R3$
PCO-CPO, our calculated values of
$E_{\text{FM}}\mathtt{-}E_{\text{AFM}}$ (per formula unit) are $-$80
meV and $-$62 meV within the LDA+\emph{U} formalism, respectively,
the corresponding bulk moduli \emph{B}=164 (\emph{B$^{\prime}$}=7.9)
and 153 GPa (\emph{B$^{\prime}$}=4.3) for AFM phase, respectively,
\emph{B}=157 (\emph{B$^{\prime}$}=5.6) and 150 GPa
(\emph{B$^{\prime}$}=4.4) for FM phase, respectively, and spin
moments of Cr ions are 2.79 and 2.92 $\mu_{B}$ for AFM
configuration, respectively. Averaged to every Cr$^{4+}$ ion, the
$E_{\text{FM}}\mathtt{-}E_{\text{AFM}}$ of $R3$ PCO-CPO is about
$-$31 meV, which is almost consistent with recent experimental
report \cite{ArevaloInorg} of the AFM ordering at
$T_{\text{N}}$$\mathtt{\sim}$245 K. For the bulk moduli, values of
$R3c$ PCO are slightly smaller than the experimental value of Phase
II \emph{B}=187 GPa \cite{Xiao}, while values of $R3$ PCO-CPO are
prominently larger than the experimental value of Phase I. The
abnormal high compressibility of Phase I is due to the fact that the
CrO$_{6/2}$ to PbO$_{6/2}$ transition has been compressed to occur
under low pressure (see below). For spin moments, results of FM
phase are almost equal to the AFM phase for PCO-CPO in $R3$ phase
and PCO in both $Pm\bar{3}m$ and $R3c$ phases.

\begin{table}[ptb]
\caption{Calculated lattice constant $a$, rhombohedral angle
$\alpha$, volume $V$, and Wyckoff parameters for PCO in $R3c$ phase
and PCO-CPO in $R3$ structure. For $R3c$ phase, the Wyckoff
positions 2$a$ ($x,x,x$) and 6$b$ ($x,y,z$) refer to the cations and
anions, respectively. In case of the $R3$ structure, the
corresponding Wyckoff labels are 1$a$ ($x,x,x$) and 3$b$ ($x,y,z$).}%
\label{lattice}%
\begin{ruledtabular}
\begin{tabular}{cccccccccccccc}
&&PCO&PCO-CPO\\
\hline
space group&&$R3c$&$R3$\\
$a$ [{\AA}]&&5.381&5.800\\
$\alpha$ [$^{\circ}$]&&60.86&55.12\\
$V$ [{\AA}$^{3}$]&&112.33&122.31\\
Pb&$x$&0.977&0.987/0.729\\
Cr&$x$&0.218&0.209/0.525\\
O&$x$&0.542&0.533/0.436\\
&$y$&0.958&0.924/0.139\\
&$z$&0.392&0.377/0.815\\
\end{tabular}
\end{ruledtabular}
\end{table}

\subsection{Pressure behaviors}
The equation of states of AFM PCO-CPO in $Pm\bar{3}m$ and $R3$
phases, AFM PCO in $Pm\bar{3}m$ and $R3c$ phases, and the
experimental measured pressure-volume data from Ref. [\cite{Xiao}]
are presented in Fig. 4. For $Pm\bar{3}m$ and $R3c$ phases of PCO
($R3$ phase of PCO-CPO), the relative smaller volumes calculated in
our scheme compared with experimental Phase II (Phase I) originates
from the typical overbinding character of LDA. From Fig. 4, one can
find that our calculated volume collapses of $Pm\bar{3}m$ PCO-CPO to
$Pm\bar{3}m$ PCO and $R3$ PCO-CPO to $R3c$ PCO at experimental phase
transition pressure 1.6 GPa is about 12.4\% and 8.0\%, respectively.
The former value is larger than the measured value (9.8\%) in recent
experiments \cite{Xiao}, while the latter value is smaller than
that. Underestimation of the volume collapse value, from $R3$
PCO-CPO to $R3c$ PCO, can be attributed to the experimental fact
that the CrO$_{6/2}$ to PbO$_{6/2}$ transition has been compressed
to occur under low pressure of around 0.1-1.6 GPa in the
experimental compound ``Phase I". This kind of partial transition
leads to abnormal high compressibility of Phase I compared with
CaCrO$_{3}$, SrCrO$_{3}$, and high-pressure Phase II of PbCrO$_{3}$
\cite{Zhou,Xiao}.

\subsection{Electronic structure}
Figure 5 shows the total density of states (DOS) as well as the
projected DOS for the Cr 3\emph{d}, Cr 4\emph{s}, and O 2\emph{p} orbitals for AFM
PCO in $Pm\bar{3}m$ and $R3c$ phases and AFM PCO-CPO in $R3$ phase
at selective values of \emph{U} within LDA+\emph{U} formalism. Corresponding band-structures calculated with \emph{U}=4 eV are presented in Fig. 6, where both spin-up and spin-down results are plotted. Since spin-down results for PCO in $Pm\bar{3}m$ and $R3c$ phases are same with their spin-up results, as indicated in Figs. 6(a)-6(b), we only plot in Figs. 5(a)-5(d) the spin-up results. For AFM PCO-CPO in $R3$ phase, slight differences between spin-up and spin-down can be seen in Figs. 5(e) and 6(c). Overall, results of PCO in both $Pm\bar{3}m$ and $R3c$ phases indicate that
the AFM PCO is metallic without accounting for or after switching on
the on-site Coulomb repulsion [see Figs. 5(a)-5(d) and 6(a)-6(b)]. This fact
conflicts with the experimental observations that the ``PbCrO$_{3}$"
is AFM semiconductor with 0.27 eV or 0.11 eV activation energy in
different temperature ranges \cite{Chamberland1,ArevaloInorg}. In
our calculations even increasing the amplitude of \emph{U} up to 8
eV, the metallic state has not changed for these two phases of PCO.
The metallic ground states have also been observed for NM and FM
phases. Besides, inclusion of the spin-orbit coupling (SOC) and
noncollinearity also can not open a gap at the Fermi level. Thus, we
only can conclude that the high-pressure phase of PbCrO$_{3}$ is a
conductor. Additionally, results of CPO and PCO-CPO in $Pm\bar{3}m$
phase also show that they are conductors. Although a gap is opened
with the Hubbard \emph{U}=6 eV for $Pm\bar{3}m$ PCO-CPO (not shown),
the calculated insulating band gap (2.12 eV) is prominently larger
than the experimental values \cite{Chamberland1,ArevaloInorg}. In
present study, we prefer to believe that the LDA+\emph{U} with
\emph{U}=4 eV can give a correct depictions of the ground state
electronic structures for PCO and PCO-CPO. For AFM PCO-CPO in $R3$
phase, our LDA+\emph{U} with \emph{U}=4 eV calculation open an
insulating band gap of about 0.48 eV [see Fig. 5(e) and 6(c)]. Figure 6(c) clearly indicates that the valence band maximum (VBM) appear at Z point and conduction
band minimum (CBM) at $\Gamma$ point in BZ. We find that both VBM and CBM have
predominant O 2\emph{p} state character mixed with significant
Cr 3\emph{d} contribution. Although the calculated value of band gap (0.48 eV) is
almost two times of the experimental value 0.27 eV
\cite{Chamberland1,ArevaloInorg}, since the activation energy of the
Phase I perovskite increases with lowing temperature
\cite{ArevaloInorg}, we believe that our calculation (valid only at
0 K) can give a proper depictions of the electronic structures for
low-pressure phase of PbCrO$_{3}$. Our calculations clearly
illustrate that the ``PbCrO$_{3}$" will occur an insulator-metal
transition together with the phase transition of $R3$ PCO-CPO to
$Pm\bar{3}m$/$R3c$ PCO upon compression.

We note that the conductivity of SrCrO$_{3}$ and CaCrO$_{3}$ exists
controversial \cite{Zhou,Ortega,Komarek,Streltsov,Lee}. Theoretical
calculation for CaCrO$_{3}$ \cite{Streltsov} predicted metallic
ground state with LDA and insulating state with LDA+\emph{U}. For
SrCrO$_{3}$ \cite{Lee}, the metallic ground sate was observed either
with LDA or LDA+\emph{U}. Besides, using LDA+\emph{U} method with
\emph{U}=4 eV, Lee \emph{et al.} \cite{Lee} successfully predicted
an orbital-ordering transition from $t_{2g}^{2}$ to
$d_{xy}^{1}(d_{xz}d_{yz})^{1}$ for SrCrO$_{3}$. No evidence of
orbital ordering within the $t_{2g}$ shell for CaCrO$_{3}$ was
observed \cite{Komarek}. In our study of PCO, CPO, and PCO-CPO
compounds, this kind of orbital ordering in Cr $t_{2g}$ states has
also not been found.

As shown in Fig. 5, inclusion of the on-site Coulomb repulsion will
lower the occupation energy of Cr 3\emph{d} electrons from the Fermi
level and elevate Cr 3\emph{d}, Cr 4\emph{s}, and O 2\emph{p} orbitals occupation levels near $-$7.0 eV to high level. As a result, localization pictures of electrons occupation appear after introducing the Coulomb repulsion. From Figs. 5(c)-5(e), one can find that the Cr 4\emph{s} contribution is limited. When Cr ions combining with O ions to form covalent/ionic bonds, part of Cr 3\emph{d} and Cr 4\emph{s} electrons will transfer to O 2\emph{p} orbital. This kind of electron transfer behavior can be read from the partial DOS pictures. Considering the effect of \emph{d}-hole creation due to 3\emph{d}-4\emph{s} hybridization, we have also examined the effect of Coulomb repulsion on Cr 4\emph{s}-shell. However, no difference of Cr 4\emph{s} orbital has been found for introducing the Coulomb repulsion into the \emph{d}-shell or the \emph{s}-shell. In Figs. 5 and 6, only \emph{d}-shell is considered to participate in the Coulomb exchange. In the whole energy domain, electronic structures of AFM PCO in $Pm\bar{3}m$ and $R3c$
phases have no evident differences. The
main occupation at the Fermi level is from Cr 3\emph{d} and O
2\emph{p} orbitals. Our results calculated at \emph{U}=0 eV are
consistent with previous calculations \cite{Jaya}. A clear
hybridization of Cr 3\emph{d} and O 2\emph{p} orbitals in the energy
range from $-$7.3 to 0.3 eV can be observed at \emph{U}=0 eV. After
switching on the \emph{U} to 4 eV, this hybridization energy range
is moved to from $-$6.2 to 0.2 eV. A well resolved peak of Cr
3\emph{d} state at around $-$0.3 eV at \emph{U}=0 eV is flatted when
the Habburd \emph{U} parameter being increased to about 4 eV. In
addition, a band gap in the conduction band is apparent under
\emph{U}=0 to 6 eV. This band gap increases from 0.6 eV at
\emph{U}=0 eV to about 2.0 eV at \emph{U}=4 eV. The main occupation
in the conduction band is from Cr 3\emph{d} orbital with some
contribution from O 2\emph{p} states. For AFM PCO-CPO in $R3$ phase,
hybridization of Cr 3\emph{d} and O 2\emph{p} orbitals in the energy
range from $-$6.3 to $-$1.0 eV is clear. One narrow peak locates
just below the Fermi level.

\section{CONCLUSIONS}
In conclusion, the ground state properties as well as the high
pressure behaviors of PCO, CPO, and PCO-CPO compounds were studied
by means of the first-principles DFT+\emph{U} method. By choosing
the Hubbard \emph{U} parameter around 4 eV within the LDA+\emph{U}
approach, FM and/or AFM ground states were achieved and our
calculated volumes, bulk moduli, spin moments, and equation of
states are in good agreement with recent experiments. While the
PCO-CPO in $R3$ phase is consistent with the experimental
low-pressure Phase I, both $Pm\bar{3}m$ and $R3c$ phases of PCO
coincide well with high-pressure Phase II. Specially, the
semiconductor nature of $R3$ PCO-CPO is in good agreement with
experiments. These observations explicitly indicate the existence of
strongly correlated electronic behaviors in these compounds. Our
electronic spectrums illustrate a clear hybridization of Cr
3\emph{d} and O 2\emph{p} orbitals in wide energy range. In contrast
to SrCrO$_{3}$, the orbital-ordering transition from $t_{2g}^{2}$ to
$d_{xy}^{1}(d_{xz}d_{yz})^{1}$ has not been found in these
materials.

\begin{acknowledgments}
We are grateful to O. Eriksson for useful discussions. This work was supported by the National Basic Research Program of
China (973 Program) (Grant No. 2010CB833102) and the National
Natural Science Foundation of China (Grant Nos. 11104170, 10974244, and
11074155).
\end{acknowledgments}

\clearpage

Figure captions:

Fig. 1: (Color online) Pictorial illustrations of (a) cubic PCO and
(b) cubic PCO-CPO in AFM configuration. For PCO-CPO, the first atom
along the [111] diagonal direction is labeled as Pb1, second Cr1,
third Cr2, and fourth Pb2. One can find that along [111] direction
in PCO the Pb and Cr alternate while in PCO-CPO the Pb and Cr
alternate in pairs. Compared with PCO, in PCO-CPO half percent of
the A site and B site atoms are exchanged while in CPO the A site
and B site atoms are totally exchanged.

Fig. 2: (Color online) Dependences of (a) total-energy differences
(per formula unit) between FM and AFM
($E_{\text{FM}}\mathtt{-}E_{\text{AFM}}$), (b) lattice parameter,
(c) bulk modulus, and (d) spin moments of Cr ions on \emph{U} for
AFM PCO in $Pm\bar{3}m$ structure.

Fig. 3: (Color online) Comparison of relative energies of two unit
cells of AFM PCO in $Pm\bar{3}m$ and $R3c$ phases, one formula unit
of AFM PCO-CPO in $Pm\bar{3}m$ and $R3$ phases, and two unit cells
of AFM CPO in $Pm\bar{3}m$ phase vs the volume. All results are
calculated within LDA+\emph{U} at \emph{U}=4 eV. Phase transitions
of $R3$ PCO-CPO to $Pm\bar{3}m$ PCO at 1.5 GPa and $R3$ PCO-CPO to
$R3c$ PCO at -6.7 GPa can be predicted by the slopes of the common
tangent rule, as shown in the inset.

Fig. 4: (Color online) The P-V relations of the AFM PCO-CPO in
$Pm\bar{3}m$ and $R3$ phases as well as AFM PCO in $Pm\bar{3}m$ and
$R3c$ phases computed in the LDA+\emph{U} formalism. Experimental
results from Ref. \cite{Xiao} are also presented. The volume
collapses at experimental phase transition pressure 1.6 GPa are
labeled.

Fig. 5: (Color online) The total DOS for AFM PCO in $Pm\bar{3}m$ and
$R3c$ phases as well as AFM PCO-CPO in $R3$ phase computed in the
LDA+\emph{U} formalism with selective values of \emph{U}. The
projected DOSs for the Cr 3\emph{d}, Cr 4\emph{s}, and O 2\emph{p} orbitals are
also shown. In panel (e), both spin-up and spin-down results are presented. The Fermi energy level is set at zero.

Fig. 6: (Color online) Band-structures of AFM PCO in $Pm\bar{3}m$ and
$R3c$ phases as well as AFM PCO-CPO in $R3$ phase computed in the
LDA+\emph{U} formalism with \emph{U}=4 eV. While the solid lines show the spin-up results, the dashed lines stand for spin-down. The Fermi energy level is set at zero as shown by the short-dashed
lines.

\begin{thebibliography}{99}
\bibitem {Zhong}W. Zhong and D. Vanderbilt,
Phys. Rev. Lett. \textbf{74}, 2587 (1995).
\bibitem {Imada}M. Imada, A. Fujimori, Y. Tokura, Rev. Mod. Phys. \textbf{70}, 1039 (1998).
\bibitem {Tokura}Y. Tokura, and N. Nagaosa, Science, \textbf{288}, 462 (2000).
\bibitem {Dagotto}E. Dagotto, Science, \textbf{309}, 257 (2005).
\bibitem {Neaton}J. B. Neaton, C. Ederer, U. V. Waghmare, N. A. Spaldin, and
K. M. Rabe, Phys. Rev. B \textbf{71}, 014113 (2005).
\bibitem {Picozzi}S. Picozzi and C. Ederer, J. Phys.: Condens. Matter
\textbf{21}, 303201 (2009).


\bibitem {Roth1}W. L. Roth and R. C. DeVries, J. Appl. Phys. \textbf{38}, 951 (1967).

\bibitem {Roth2}R. C. DeVries and W. L. Roth, J. Am. Ceram. Soc. \textbf{51}, 72 (1968).

\bibitem {Chamberland1}B. L. Chamberland and C. W. Moeller, J. Solid State Chem. \textbf{5}, 39
(1972).

\bibitem {Chamberland2}B. L. Chamberland, Solid State Commun. \textbf{5}, 663 (1967).
\bibitem {Goodenough}J. B. Goodenough, J. M. Longo, and J. A. Kafalas, Mater. Res. Bull. \textbf{3}, 471
(1968).
\bibitem {Weiher}J. F. Weiher, B. L. Chamberland, and J. L. Gillson, J.
Solid State Chem. 3, 529 (1971).

\bibitem {ArevaloJSSC}$\mathrm{{\acute{A}}}$. M. Ar$\mathrm{{\acute{e}}}%
$valo-L$\mathrm{{\acute{o}}}$pez and M. $\mathrm{{\acute{A}}}$. Alario-Franco,
J. Solid State Chem. \textbf{180}, 3271 (2007).

\bibitem {ArevaloJPCM}$\mathrm{{\acute{A}}}$. M. Ar$\mathrm{{\acute{e}}}%
$valo-L$\mathrm{{\acute{o}}}$pez, E. Castillo-Mart$\mathrm{{\acute{i}}}$nez,
and M. $\mathrm{{\acute{A}}}$. Alario-Franco, J. Phys.: Condens. Matter
\textbf{20}, 505207 (2008).

\bibitem {ArevaloInorg}$\mathrm{{\acute{A}}}$. M. Ar$\mathrm{{\acute{e}}}%
$valo-L$\mathrm{{\acute{o}}}$pez, A. J. D. santos-Garc$\mathrm{{\acute{i}}}$a,
and M. $\mathrm{{\acute{A}}}$. Alario-Franco, Inog. Chem. \textbf{48}, 5434 (2009).

\bibitem {Xiao}W. Xiao, D. Tan, X. Xiong, J. Liu, and J. Xu, PNAS
\textbf{107}, 14026 (2010).

\bibitem {Zhou}J.-S. Zhou, C.-Q. Jin, Y.-W. Long, L.-X. Yang, and J. B. Goodenough,
Phys. Rev. Lett. \textbf{96}, 046408 (2006).

\bibitem {Ortega}L. O. SanMartin, A. J. Williams, J. Rodgers, J. P. Attfield, G.
Heymann, and H. Huppertz, Phys. Rev. Lett. \textbf{99}, 255701
(2007).

\bibitem {Komarek}A. C. Komarek, S. V. Streltsov, M. Isobe, T. Moller, M. Hoelzel, A.
Senyshyn, D. Trots, M. T. Fern\'{a}ndez-D\'{i}az, T. Hansen, H.
Gotou, T. Yagi, Y. Ueda, V. I. Anisimov, M. Gruninger, D. I.
Khomskii, and M. Braden, Phys. Rev. Lett. \textbf{101}, 167204
(2008).

\bibitem {Streltsov}S. V. Streltsov, M. A. Korotin, V. I. Anisimov, and D. I. Khomskii,
Phys. Rev. B \textbf{78}, 054425 (2008).

\bibitem {Lee}K. W. Lee and W. E. Pickett, Phys. Rev. B \textbf{80}, 125133 (2009).

\bibitem {Dudarev}S. L. Dudarev, G. A. Botton, S. Y. Savrasov, C. J.
Humphreys, and A. P. Sutton, Phys. Rev. B \textbf{57}, 1505 (1998).

\bibitem {PAW}P. E. Bl\"{o}chl, Phys. Rev. B \textbf{50}, 17953 (1994).

\bibitem {Kresse3}G. Kresse and J. Furthm\"{u}ller, Phys. Rev. B \textbf{54},
11169 (1996).

\bibitem {LDA}W. Kohn and L. J. Sham, Phys. Rev. \textbf{140}, A1133 (1965).

\bibitem {GGA}J. P. Perdew, K. Burke, and Y. Wang, Phys. Rev. B \textbf{54},
16533 (1996).

\bibitem {Monk}H. J. Monkhorst and J. D. Pack, Phys. Rev. B \textbf{13}, 5188 (1972).

\bibitem {Birch}F. Birch, Phys. Rev. \textbf{71}, 809 (1947).

\bibitem {Jaya}S. Mathi Jaya, R. Jagadish, R. S. Rao, and R. Asokamani, Mod. Phys.
Lett. B \textbf{6}, 103 (1992).

\bibitem {Dedkov}Yu. S. Dedkov, A. S. Vinogradov, M. Fonin, C. K\"{o}nig, D. V.
Vyalikh, A. B. Preobrajenski, S. A. Krasnikov, E. Yu. Kleimenov, M.
A. Nesterov, U. R\"{u}diger, S. L. Molodtsov, and G. G\"{u}ntherodt,
Phys. Rev. B \textbf{72}, 060401(R) (2005); D. J. Huang, H.-T. Jeng,
C. F. Chang, G. Y. Guo, J. Chen, W. P. Wu, S. C. Chung, S. G. Shyu,
C. C. Wu, H.-J. Lin, and C. T. Chen, \emph{ibid}. \textbf{66},
174440 (2002); M. A. Korotin, V. I. Anisimov, D. I. Khomskii, and G.
A. Sawatzky, Phys. Rev. Lett. \textbf{80}, 4305 (1998); T. Tsujioka,
T. Mizokawa, J. Okamoto, A. Fujimori, M. Nohara, H. Takagi, K.
Yamaura, and M. Takano, Phys. Rev. B \textbf{56}, R15509 (1997).

\end{thebibliography}
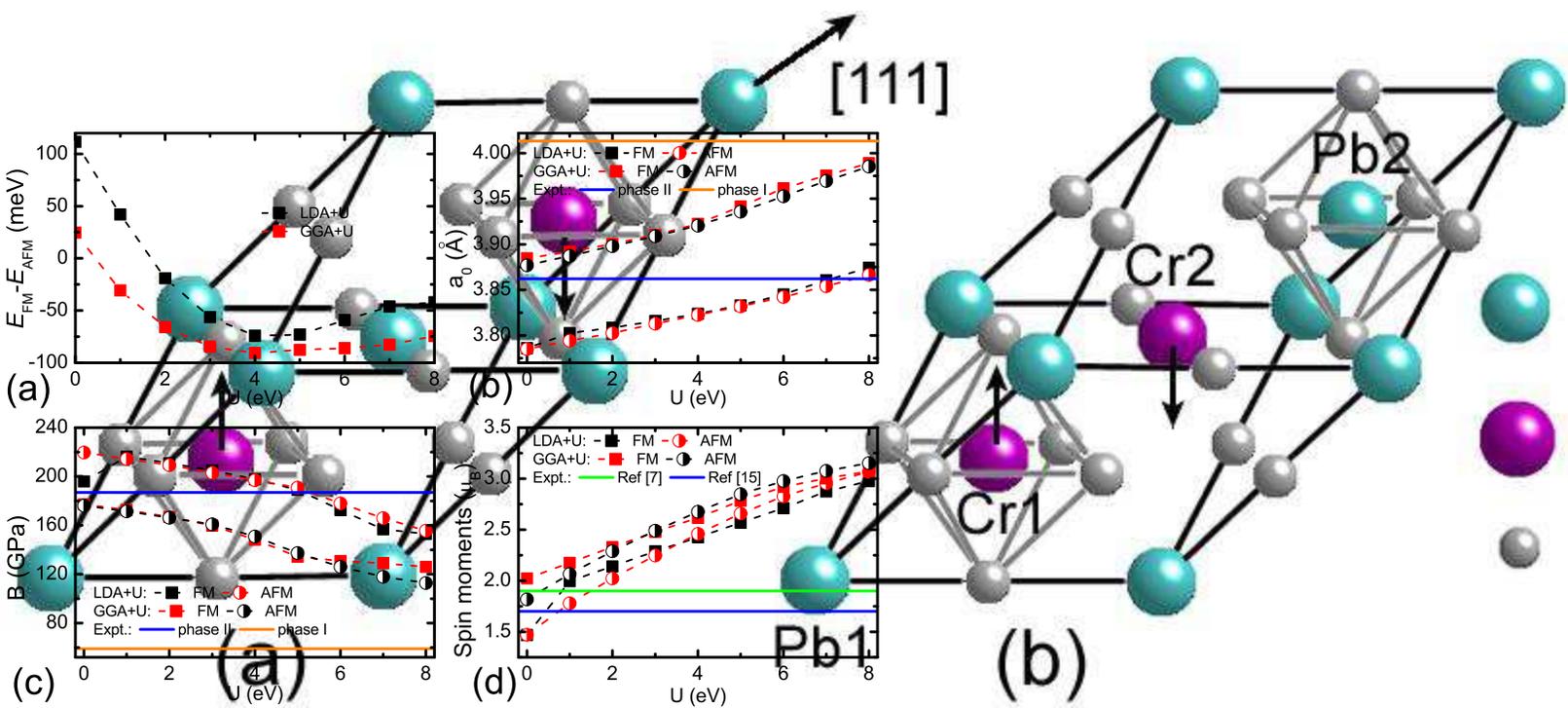
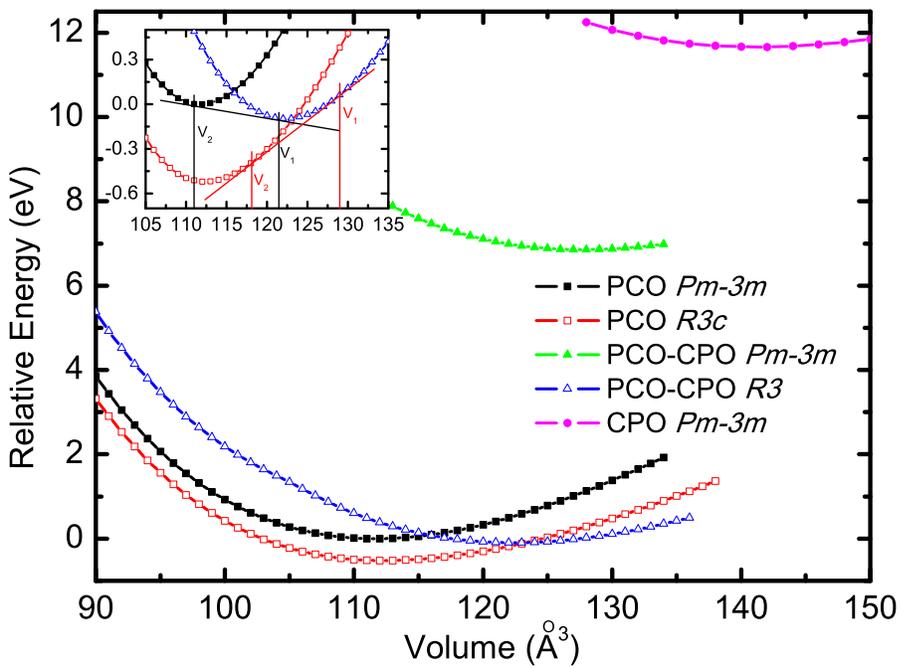
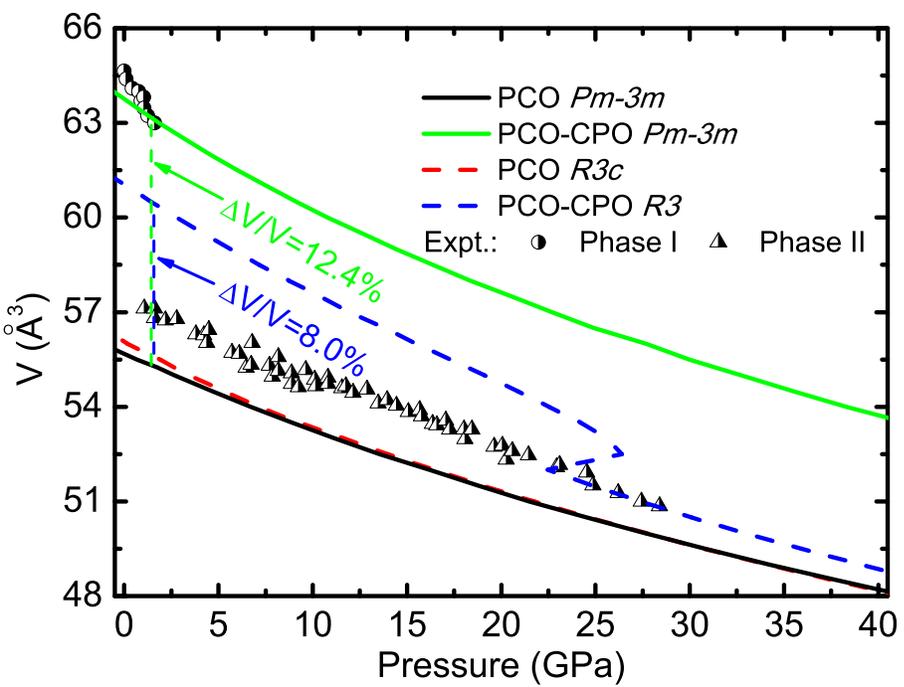
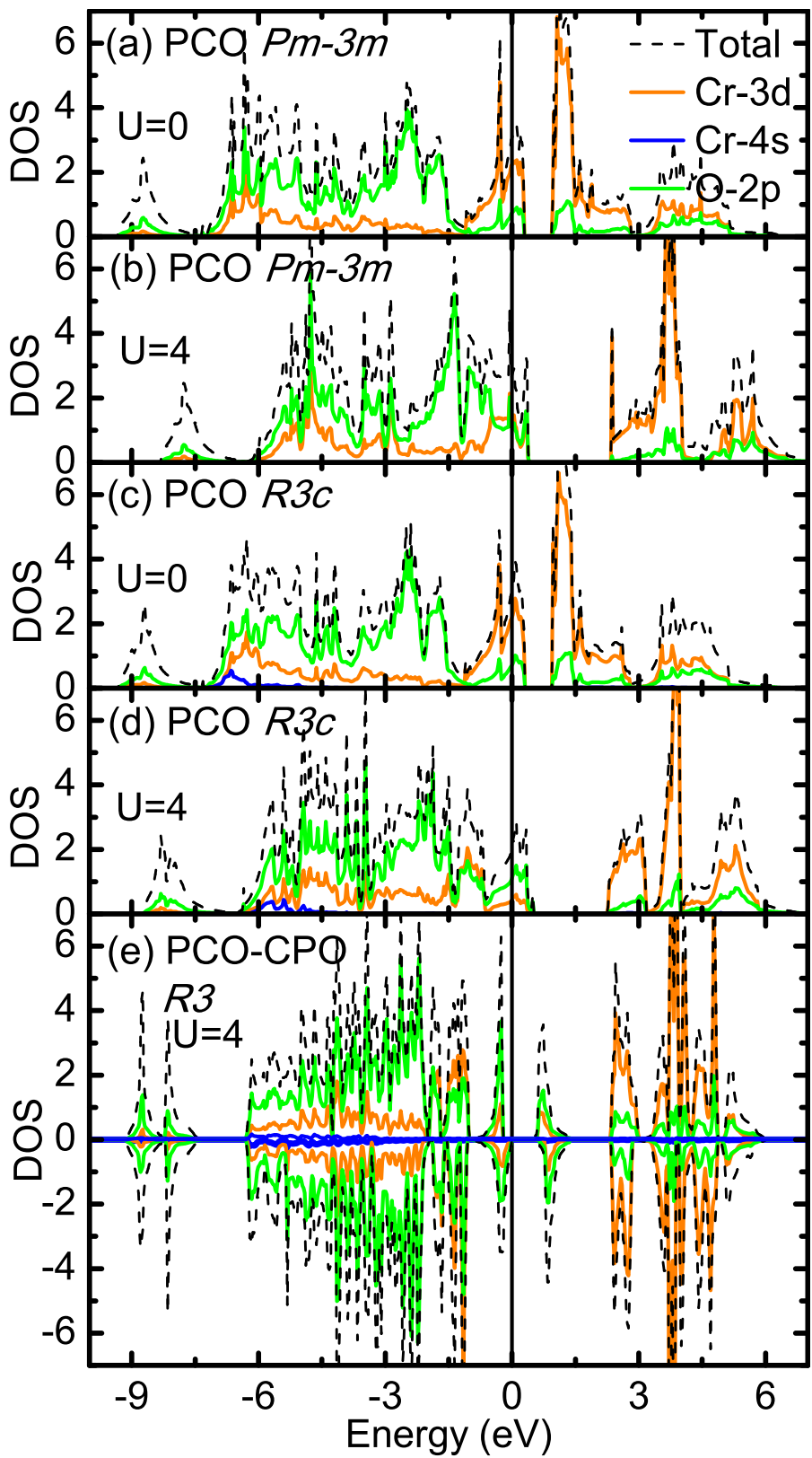
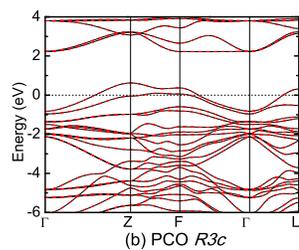
\end{document}